\documentclass[12pt,english]{article}
\usepackage[T1]{fontenc}
\usepackage[latin9]{inputenc}
\usepackage[letterpaper]{geometry}
\usepackage{array}
\usepackage{graphicx}

\makeatletter

\providecommand{\tabularnewline}{\\}

\usepackage{times,graphics,natbib}
\renewcommand{\baselinestretch}{1.5}
\usepackage{lineno}

\newcommand{\bld}[1]{\mbox{\boldmath $#1$}}

\newcommand{\smttt}[1]{\mbox{\scriptsize{#1}}}
\newcommand{\degr}{^{\circ}}

\newcommand{\matern}{Mat\'{e}rn }

\makeatother

\usepackage{babel}
\begin{document}

\title{Combining spatial information sources while accounting for systematic
errors in proxies}

\author{Christopher J. Paciorek\\
Department of Biostatistics, Harvard School of Public Health, Boston,
USA, and \\
Department of Statistics, University of California, Berkeley, USA}

\maketitle

\begin{abstract}
Environmental research increasingly uses high-dimensional remote sensing
and numerical model output to help fill space-time gaps between traditional
observations. Such output is often a noisy proxy for the process of
interest. Thus one needs to separate and assess the signal and noise
(often called discrepancy) in the proxy given complicated spatio-temporal
dependencies. Here I extend a popular two-likelihood hierarchical
model using a more flexible representation for the discrepancy. I
employ the little-used Markov random field approximation to a thin
plate spline, which can capture small-scale discrepancy in a computationally
efficient manner while better modeling smooth processes than standard
conditional auto-regressive models. The increased flexibility reduces
identifiability, but the lack of identifiability is inherent in the
scientific context. I model particulate matter air pollution using
satellite aerosol and atmospheric model output proxies. The estimated
discrepancies occur at a variety of spatial scales, with small-scale
discrepancy particularly important. The examples indicate little predictive
improvement over modeling the observations alone. Similarly, in simulations
with an informative proxy, the presence of discrepancy and resulting
identifiability issues prevent improvement in prediction. The results
highlight but do not resolve the critical question of how best to
use proxy information while minimizing the potential for proxy-induced
error.  
\end{abstract}

\section{Introduction\label{sec:IntroductionBcs}}

There has been substantial interest recently in combining observations
at spatial point locations with proxy information from remote sensing
and numerical model output, particularly in the area of air quality.
Building on \citet{Fuen:Raft:2005}, statisticians have proposed a
number of modeling approaches, often termed 'data fusion'. Most use
Bayesian hierarchical spatial models to combine information sources
with goals such as air quality management, forecasting, and exposure
prediction for health analysis. Critically, numerical models and remote
sensing retrievals often produce highly spatially-correlated surfaces,
but some of this structure may represent spatially-correlated error
with regard to the quantity of interest. For example, a numerical
model may overpredict a pollutant over a wide area because of shortcomings
in information on emissions sources, while cloud or surface contamination
may cause spatially-correlated errors in satellite retrievals. This
'error' is better termed discrepancy when the proxy is not designed
to estimate the focal process of interest but rather some related
quantity. 

In statistical formulations of the problem, the possibility of proxy
discrepancy has generally been acknowledged, but previous modeling
efforts have often placed strong constraints on the structure of the
discrepancy for reasons of identifiability or computational feasibility.
\citet{Fuen:Raft:2005} proposed the following general model with
their proxy (numerical model output) treated as data via a second
likelihood,
\begin{eqnarray}
Y_{i} & = & L(\bld{s}_{i})+\epsilon_{i}\nonumber \\
A_{j} & = & \phi(\bld{s}_{j})+\beta_{1}(\bld{s}_{j})L(\bld{s}_{j})+e_{j}.\label{eq:fuentesModel}
\end{eqnarray}
The model relates both the gold-standard observations, $Y_{i}$, $i=1,\ldots,n$,
and the proxy values, $A_{j}$, $j=1,\ldots,m$ ($A$ for auxiliary),
to the latent, true process of interest (the focal process), $L(\bld{s})$,
at location $\bld{s}$. For the moment I suppress any change-of-support
manipulations in defining $L(s_{j})$ when $s_{j}$ is an area. $\phi(\cdot)$
and $\beta_{1}(\cdot)$ are additive and multiplicative bias terms.
In general it will be difficult to identify both $\phi(\cdot)$ and
$\beta_{1}(\cdot)$, and \citet{Fuen:Raft:2005} use a scalar $\beta_{1}$
and take $\phi(\cdot)$ to be a simple polynomial in the spatial coordinates.
Critically, since the additive bias has a very low dimensional representation,
this approach assumes that all of the small- and moderate-scale spatial
structure in the proxy is signal with respect to $L(\cdot)$. \citet{Paci:Liu:2009}
used such a structure (\ref{eq:fuentesModel}), with remote sensing
retrievals playing the role of the proxy, but chose a reduced rank
spline basis for additional flexibility in modeling the discrepancy.
We found that model fitting was sensitive to the number of basis functions,
with increasingly better fits as the number of basis functions increased,
such that computational complexity prevented fitting a model with
sufficiently many basis functions to model $\phi(\cdot)$ adequately.
Other recent work has used such moderately flexible specifications
for quantities analogous to $\phi(\cdot)$: \citet{Fuen:etal:2008}
used a small number of basis functions, while \citet{McMi:etal:2010}
used b-splines in two dimensions. 

The danger in limiting the flexibility of the discrepancy representation
is that systematic discrepancies will bias prediction of the spatial
process of interest in subdomains and that correlated uncertainty
will not be properly acknowledged. In short, spatially-correlated
discrepancy in the proxy may look like signal because of the dependence
structure but will cause spurious features in the predictions. Overly
constrained discrepancy terms implicitly assume the proxy is useful
and data fusion successful, but changes in prediction from including
the proxy may primarily reflect discrepancy. Gold-standard data can
help to assess the potential for discrepancy at scales resolved by
the data. For large scales (relative to the data density), one can
hope to estimate and adjust for the discrepancy using the data. At
smaller scales, one can at best hope to discount proxy information
if the data indicate discrepancy. The key to this effort lies in using
a sufficiently flexible model specification for the spatial discrepancy.

I propose to represent the discrepancy, $\phi(\cdot)$, using a computationally-efficient
Markov random field (MRF) specification that is sufficiently flexible
to model discrepancies at a variety of spatial scales. This MRF specification
(Rue and Held, 2005, Sec. 3.4.2; Yue and Speckman 2010) approximates
a thin plate spline while retaining the sparse precision matrix structure
of more widely used MRFs, such as conditional autoregressive (CAR)
models based on neighborhood adjacencies. For proxy variables, which
are often very high-dimensional, modeling $\phi(\cdot)$ efficiently
is critical. The ability of the proposed MRF specification to capture
variation at a range of spatial scales stands in contrast to standard
spatial representations: (1) reduced rank basis function approaches
\citep{Kamm:Wand:2003,Rupp:etal:2003,Bane:etal:2008} omit smaller-scale
structure to improve computational efficiency, while (2) traditional
CAR models are computationally tractable and capture small-scale variability,
but estimate small-scale variation even when it is not present, as
discussed in Section \ref{sub:Markov-random-field}. 

The advantage of this approach is that it allows for the real possibility
that discrepancy can occur at a variety of scales, in particular at
smaller scales than have been possible in other analyses. In many
applications, we have no reason to think that the proxy is only biased
at larger scales, so a realistic statistical model should allow for
spatially-correlated discrepancy at smaller scales. Unfortunately,
the latent process of interest and the discrepancy may occur at similar
scales. Flexible representation of the discrepancy may result in poor
identifiability as the model attempts to decompose variability in
the proxy between signal, $L(\cdot)$, and noise, $\phi(\cdot)$.
The open question is whether the statistical model can disentangle
signal from noise in the proxy data source sufficiently to improve
prediction of the process of interest. Note that the lack of identifiability
is inherent in the scientific context and is reflected in the proposed
model, which does not impose artificial constraints to improve identifiability. 

The potential for improving prediction using data fusion is very appealing
in applications. However, \citet{Paci:Liu:2009} found no improvement
in fine particulate matter (PM) prediction using two satellite aerosol
products. Compared to simply kriging the monitor values, \citet{McMi:etal:2010}
found no improvement in mean square prediction error of PM using output
from the Community Multiscale Air Quality (CMAQ) atmospheric chemistry
model in a Bayesian hierarchical model but did find that the fusion
improved bias and coverage results. \citet{Sahu:etal:2009} found
a statistically significant regression coefficient for the relationship
between CMAQ output and their process of interest, but the magnitude
of the coefficient was small, and there appears to be little evidence
of predictive improvement from including the proxy. \citet{Berr:etal:2010}
did find improvement in prediction of ozone relative to ordinary kriging
without covariates when including the CMAQ proxy as a covariate.

The contributions of this work are two-fold. First, I present a methodological
approach that focuses attention on the following questions: is the
proxy useful and data fusion successful; at what scales is the proxy
useful and at what scales does it distort predictions; and can one
sufficiently distinguish proxy signal and noise? I propose a specific
diagnostic for assessing the scales of discrepancy as a standard tool
in data fusion. I use the methodology to assess the use of proxies
for monthly average PM in the eastern United States. Second, given
the identifiability issues of the problem, I highlight the importance
and implications of the scales and spatial structure of the discrepancy
and show that the lack of identifiability makes it difficult in some
circumstances to improve spatial predictions using proxy information.

\section{Air pollution example}

The specific scientific goal is to improve spatio-temporal predictions
of fine PM (also called $\mbox{PM}_{2.5}$, and henceforth simply
PM) relative to current statistical modeling efforts that combine
smoothing with land use and meteorological covariates \citep{Yano:etal:2009,Paci:etal:2009,Szpi:etal:2010}.
In recent years, researchers have considered using both remotely-sensed
aerosol optical depth \citep{Paci:Liu:2009,vanD:etal:2010} and deterministic
model output \citep{Sahu:etal:2009,McMi:etal:2010} as proxies for
PM. Rather than introducing the complexity of a full spatio-temporal
model, I analyze individual months separately here, but \citet{Paci:Liu:2011}
present a spatio-temporal extension of this work. Note that \citet{Paci:etal:2009}
found little improvement from accounting for temporal correlation
when interest focused on monthly average air pollution exposure.  Of
course the relatively usefulness of a proxy may change with temporal
scale, as does the importance of temporal correlation, so results
may differ when averaging over shorter periods.

The gold-standard observations are monthly averages of available 24-h
fine PM concentrations in the eastern U.S. from the U.S. EPA Air Quality
System. The first analysis involves the spatial variation for each
month of 2004 for the mid-Atlantic region with aerosol optical depth
(AOD) from the moderate resolution imaging spectroradiometer (MODIS)
instrument as the proxy. Each orbit of the satellite provides a snapshot
of a swath of the eastern U.S. at approximately 10:30 am, which produces
an AOD value for any fixed location every 1-2 days. AOD is a dimensionless
measure of light extinction through the entire vertical column of
the atmosphere that is generally correlated with PM at the ground
surface, with a correlation of monthly averages here of 0.58. MODIS
AOD is estimated by an algorithm based on light sensed by the satellite
instrument and is available on an irregular grid with an approximate
resolution of 10 km. 

The second analysis explores the usefulness of a proxy over a larger
spatial domain, considering spatial variation for each month of 2001
for the entire eastern U.S. with CMAQ output from a 36 km-resolution
model run as the proxy. Hourly CMAQ-estimated PM is available on a
regular 36 km grid at 14 vertical levels from this run; I use the
24-h averages from the lowest level to best match PM at the ground
surface, which produces a correlation of monthly averages of 0.52.
CMAQ relies on meteorological and emissions inventory inputs, and
was designed for short-term air quality applications, so there has
been limited evaluation of long-term average output. One concern with
the first analysis is that some PM monitors report only every third
or sixth day, which increases the noise in the PM monthly averages.
To avoid such temporal misalignment, I restricted the second analysis
to monthly averages of PM data and CMAQ output from every third day,
using only ground monitors reporting every day or every third day.
Full details on the various data sources and data pre-processing manipulations
for both analyses are provided in our peer-reviewed Health Effects
Institute report \citep{Paci:Liu:2011}.

Figs. \ref{fig:predPlots1} and \ref{fig:predPlots3} show examples
of raw PM observations (top rows) and raw proxy data (bottom rows)
used in the analyses.

\section{Model and methods}

In this section I outline the basic modeling approach, with technical
details of the two specific models used in the examples and the Markov
chain Monte Carlo (MCMC) fitting methods provided in the Appendix.

\subsection{Spatial latent variable model \label{sub:Spatial-latent-variable}}

I propose the following basic spatial model 
\begin{eqnarray}
Y_{i} & = & \beta_{y}(x_{y,i})+\bld{P}_{Y,i}\bld{L}+\epsilon_{i}\nonumber \\
A_{j} & = & \phi(s_{j})+\beta_{1}\bld{P}_{A,j}\bld{L}+e_{j}\label{eq:coreModel}
\end{eqnarray}
where notation follows (\ref{eq:fuentesModel}), with $\bld{L}$ the
vectorized representation of a spatial latent variable represented
the process of interest on a fine base grid. $\beta_{y}(x_{y,i})$
is an regression function that represents sub-grid scale variability
based on a single (for notational simplicity) covariate, while $\bld{P}_{Y,i}$
and $\bld{P}_{A,j}$ are rows of mapping matrices that pick off elements
of $\bld{L}$ to map to the observations and proxy values respectively.
$\bld{P}_{A}$ will generally also weight the focal process values
to account for spatial misalignment of the proxy and base grids (particularly
relevant for irregular remote sensing grids). By virtue of relating
point-level measurements to a grid-based latent process, the model
has some of the flavor (and the computational advantages) of the measurement
error model of \citet{Sahu:etal:2009}, although $\beta_{y}(\cdot)$
allows for sub-grid heterogeneity. I take the error terms, $\epsilon_{i}$
and $e_{j}$, to be normally-distributed white noise.

I then represent $L(\bld{x},\bld{s})=\sum_{p}\beta_{L,p}(x_{L,p})+g(\bld{s})$
as the sum of multiple regression terms, $\beta_{L,p}(x_{L,p})$,
where $x_{L,p}$ is the $p$th covariate, and remaining spatial variation,
$g(\bld{s})$. $\beta_{y}(\cdot)$ and $\beta_{L,p}(\cdot)$ may be
simple linear terms or smooth regression terms to capture nonlinearity,
in which case I use the mixed model formulation of a penalized thin
plate spline (cubic radial basis functions in one dimension), where
the basis coefficients' variance component controls the amount of
smoothing \citep{Rupp:etal:2003,Crai:etal:2005b}. When the covariates
are able to represent most of the small-scale variation in the focal
process, $g(\cdot)$ need only explain large-scale variation, so one
approach is represent $g(\cdot)$ using a penalized thin plate spline,
as I do for the analysis in the mid-Atlantic region (Section \ref{sub:aodResults}).
Alternatively, when the knot-based representation of $g(\cdot)$ requires
so many knots that computations bog down, either because of small-scale
process variation or a large domain, $g(\cdot)$ can be represented
as a MRF, as described next for $\phi(\cdot)$ and used in the eastern
U.S. analysis (Section \ref{sub:subnatl}).

For $\phi(\cdot),$ I use the MRF approximation of a thin plate spline,
$\bld{\phi}\sim\mathcal{N}_{m-3}(0,(\kappa\bld{Q})^{-})$, described
in detail in Section \ref{sub:Markov-random-field}. $\bld{Q}$ is
the MRF weight matrix, with rank $m-3$, hence the use of the generalized
inverse, while $\kappa$ is a precision parameter. This approach allows
the discrepancy term to represent either smooth large-scale variation
or wiggly small-scale variation depending on the data, while providing
computational feasibility. I work with $m=17500$ and $m=5621$ pixels
in the two analyses. Note that integrating over $\bld{\phi}$ produces
a spatially-correlated proxy covariance structure. Given the equivalence
between a stochastic process in the mean and integrating over that
process to move the variation into the covariance, I believe the distinction
between representation in the mean and variance is artificial \citep[p. 114]{Cres:1993}
and instead focus on understanding the scale of the discrepancy.

\subsection{Markov random field specification\label{sub:Markov-random-field}}

MRF models, in particular the standard CAR model, often use simple
binary weights in which direct neighbors are given a weight of one
and all other locations are given zero weight. However, such models
have realizations with unappealing properties. \citet{Besa:Mond:2005}
show that the model is asymptotically equivalent (as the grid resolution
increases) to two-dimensional Brownian motion (the de Wijs process)
and \citet{Lind:etal:2011} show that the model approximates a GP
in the \matern class with the spatial range parameter (in distance
units) going to infinity and the differentiability parameter going
to zero. Brownian motion has continuous but not differentiable sample
paths, so it is not surprising that the process realizations of standard
CAR models are locally heterogeneous, as seen next, regardless of
the value of the process' variance component.

\begin{figure}
\includegraphics[scale=0.8]{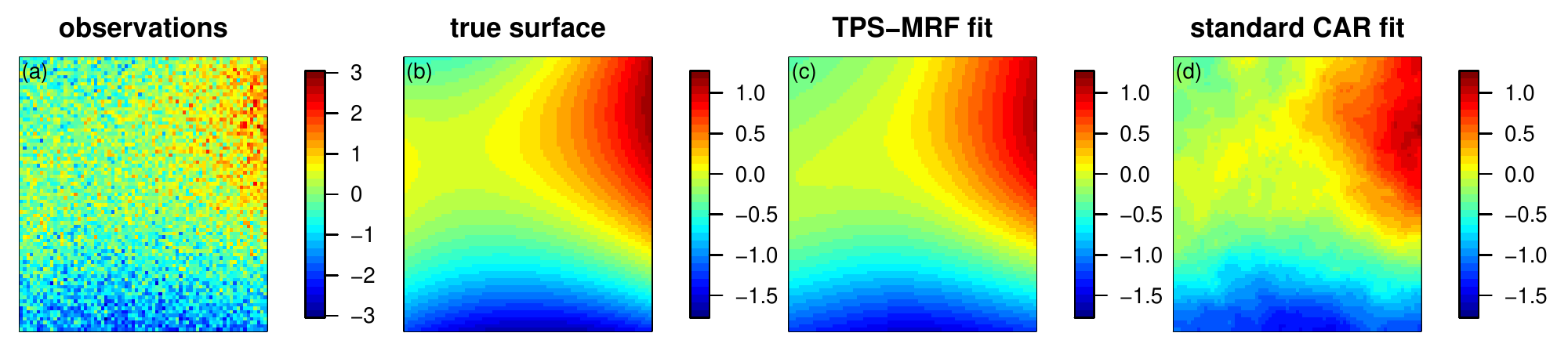}

\caption{For simulated data (a) based on white noise added to a smooth true
surface (b), (c) shows the posterior mean under the TPS-MRF model
and (d) the posterior mean under a MRF model with a standard CAR neighborhood
structure. \label{fig:fittedMRF}}
\end{figure}

A more flexible alternative that has received little attention is
an intrinsic Gaussian MRF whose weight structure is motivated by the
smoothness penalty that induces the thin plate spline (henceforth
denoted TPS-MRF) \citep{Rue:Held:2005,Yue:Spec:2010}. For a regular
grid, the TPS-MRF extends the neighborhood further from the focal
cell and introduces negative weights (details on the exact weights,
including boundary effects, can be found in \citet[Appendix C]{Paci:Liu:2011}).
Fig.\ref{fig:fittedMRF} shows the fitted posterior mean surface under
the standard CAR and TPS-MRF models for a simulated dataset whose
true surface is very smooth; note the local heterogeneity induced
by using the CAR model. The TPS-MRF can also represent fine-scale
(albeit differentiable) variability (not shown), in the following
sense. If the spatial surface varies at fine scales, the TPS-MRF can
capture this variation given sufficient data, and in doing so it will
also follow any larger scale variations. A referee suggested the alternative
of more explicitly decomposing scales by combining the standard CAR
model to represent small-scale variation with basis functions such
as regression splines or reduced rank kriging to capture large-scale
variation. 

\subsection{Spatial scales of discrepancy}

\subsubsection{Discrepancy scenarios and identifiability\label{sub:Discrepancy-scenarios}}

One important goal of including the spatial discrepancy term is to
understand the scales at which the proxy and the process of interest
are well- and poorly-correlated. I posit a range of potential relationships
and the implications for making use of proxy signal, as well as the
implications of overly constraining the discrepancy process. These
represent extreme scenarios, so real applications will likely involve
a combination of scenarios.
\begin{itemize}
\item White noise discrepancy: the spatial structure in the proxy mirrors
the focal process but there is fine-scale discrepancy at the scale
of pixels that can be treated as white noise. Under this scenario,
there is no need for $\phi(\cdot)$ given the white noise error structure,
$\{e_{j}\}$. Smoothing over the proxy gives information about the
process of interest.
\item Small-scale discrepancy: the proxy accurately reflects the focal process
at large scales, but there is smaller-scale spatially-correlated discrepancy.
Under this scenario, models without a sufficiently flexible discrepancy
term may treat the discrepancy as signal since it is not white noise,
particularly when the proxy sample size overwhelms the gold standard
observations. This allows the variation in the proxy to be explained
by a spatial process rather than by white noise (with higher accompanying
posterior density, analogous to a penalized likelihood setting), even
at the expense of increasing the observation error variance. In contrast,
with a flexible representation of $\phi(\cdot)$, the model can treat
the discrepancy as having small-scale spatial dependence, thereby
ignoring this component of variation in the proxy. \\
Note that without dense gold standard data, the small-scale signal
and noise in the proxy are not identifiable, but improved small-scale
prediction may be achievable by attributing some of the small-scale
variability in the proxy to the focal process. Estimation of $\beta_{1}$
thus involves a bias-variance tradeoff. Finally, unless there are
large spatial gaps in the observations, it's not clear if using the
proxy to help estimate large-scale variation will improve upon simply
smoothing the gold standard data.
\item Large-scale discrepancy: the proxy accurately reflects small-scale
variation in the focal process, but there is large-scale discrepancy.
In this case, one can correct for discrepancy through $\phi(\cdot)$
(needing only moderate amounts of gold standard data) and predict
small-scale variation in the focal process from small-scale variation
in the proxy. In this case it would be appropriate to constrain $\phi(\cdot)$
to vary only at larger spatial scales. \\
One potential difficulty is that discrepancy occurring at similar
scales as the focal process can cause problems with identifiability.
The attribution of variability between the discrepancy and the focal
process is determined through a complicated tradeoff between the likelihood
for the data, the likelihood for the proxy, and penalization of complexity
via the spatial process priors. Unless one removes the large-scale
variation in the proxy via a projection, discrepancy can 'leak' into
the focal process predictions, analogous to the bias occurring in
the partial spline setting \citep{Rice:1986}.
\item Uninformative proxy: the proxy and focal process are at best weakly
related at all scales. A model without a flexible discrepancy term
may have trouble representing the proxy reasonably, with focal process
predictions largely reflecting the proxy when the proxy sample size
overwhelms the gold standard data as in the small-scale discrepancy
scenario. 
\end{itemize}

\subsubsection{Spatial discrepancy diagnostic\label{sub:Spatial-discrepancy-diagnostic}}

To assess the spatial scales of the discrepancy term, I propose to
use a spatial variogram-based diagnostic introduced by \citet{Jun:Stei:2004}
for assessing numerical model performance relative to observations.
They calculate variograms for the model output, observations, and
model error (defined as model output minus observations). They propose
the ratio of the model error variogram to the sums of the variograms
for model output and observations as a distance-varying diagnostic
of the spatial variation in the observations captured by the model
output. If the model output and observations were independent, then
the variogram of the error would be the sum of the variograms of output
and observations, so the ratio would be one on average. My analog
plugs the analogous quantities from the statistical model (\ref{eq:coreModel})
into their diagnostic, using $\beta_{1}\bld{L}$ as the best estimate
of the focal process, scaled to the units of the proxy: 
\[
M(d)=\frac{\mbox{Variog}(\bld{\phi};d)}{\mbox{Variog}(\bld{\phi}+\beta_{1}\bld{L};d)+\mbox{Variog}(\beta_{1}\bld{L};d)}.
\]
$M(d)$ is interpreted as the proportion of the variation in the proxy
that is accounted for by the discrepancy term, as a function of spatial
scale, quantified by distance, $d$. The diagnostic has the following
appealing extremal properties: when $\beta_{1}=0$ or if $\bld{\phi}$
offsets all the variation in the focal process at a given scale, $M(d)=1$,
indicating all of the variability in the proxy is explained by discrepancy.
In the ideal situation when $\bld{\phi}=\bld{0}$ in general or has
no variation at a given scale, then $M(d)=0$, indicating all of the
variability in the proxy at the scale is explained by the process
of interest.

\subsection{Marginalization and MCMC sampling}

The proposed model contains two latent processes that can trade off
in explaining variation in the proxy. In addition there is cross-level
dependence between each latent process and its hyperparameters. In
such situations, marginalization over the process or joint sampling
of a process and its hyperparameters \citep{Rue:Held:2005} are often
used, when possible, to improve MCMC convergence and mixing. Here
the high-dimensionality of the processes complicates matters further.
The modeling approach allows for marginalization over the process
values and then efficient computations based on sparse matrix manipulations,
with details given in the Appendix. I marginalize first over the MRF(s)
and then over any basis coefficients to produce a representation of
the marginal posterior that can be calculated based on sparse matrix
manipulations. Note that even with these efforts, sampling can be
slow; for the mid-Atlantic and eastern U.S. analyses run times were
on the order of 6 hours and 3-5 days, respectively. Much of this was
required to achieve good mixing of the variance components for the
MRF and spline basis coefficients without the use of informative priors.

\section{Simulations}

\subsection{Methods}

To assess the performance of the approach I fit a simplified model
to simulated data under a variety of scenarios:
\begin{enumerate}
\item Large- and small-scale discrepancy present with $\beta_{1}=1$
\item Large- and small-scale discrepancy with $\beta_{1}=0$
\item No spatial discrepancy, $\beta_{1}=1$
\item Large-scale discrepancy only, $\beta_{1}=1$
\item Small-scale discrepancy only, $\beta_{1}=1$
\item Scenario 1 but with sparse observations ($n=40$)
\end{enumerate}
I constructed the simulation to mimic the mid-Atlantic analysis, with
similar number of gold-standard observations, $n=171$ (except for
scenario 6) and including a subset of the covariates used there. The
proxy is fully-observed over the entire 175x100 grid with no misalignment
and with white noise discrepancy in all cases, but I calculate the
likelihood only for the 15,157 land-based pixels. I used a GP with
\matern covariance, parameterized as in \citet{Bane:etal:2004},
with $\nu=2$ and varying values of $\rho=1/\phi$ to simulate the
residual spatial surface, $\bld{g}$ (effective range: 340 grid cells),
and (independently) the discrepancy, $\bld{\phi}$, operating at two
different scales (effective ranges: 413 and 24 grid cells) (when
not zeroed out in a given scenario). Note that the surface generation
model differs from the spatial process representations in the model.
I included population density, elevation, road density in the two
largest road size classes, and county average emissions as linear
(on the log scale for the first two terms and square root scale for
the other three) covariates for the focal process with coefficients
of $0.2$, $-0.3$, $0.01$, $0.01$, and $0.1$. When fitting the
model, I included road density in the third largest road size class
rather than the first two largest to introduce some model misspecification.
In addition, I introduced mismatch in some of the transformations
between data generation and model fitting by using a linear function
of elevation, truncated to a maximum of 500 m, a linear function of
the road density, and the log transform of county emissions. As in
the real analyses, I included local covariates to capture within-pixel
variability. I generated local variability as a function of distances
(m), $d_{1}$ and $d_{2}$, to the nearest major road in two size
classes as $50d_{1}^{-0.77}+10d_{2}^{-0.77}$ . To fit the model I
used linear terms of these distances, truncated at 10 and 500 m. 

I carried out 10 replications, resampling the spatial surfaces and
error terms. The variances of the different components were $0.55^{2}$
and $1.73^{2}$ for the normally-distributed proxy and observation
white noise, respectively, $0.84^{2}$ for the empirical within-pixel
variability from the two local covariates combined, $0.93^{2}$ for
the empirical variability in the true process from the five land use
covariates, $2.5^{2}$ for the residual spatial surface component
of the true process, $1.64^{2}$ for the small-scale discrepancy,
and $2.0^{2}$ for the large-scale discrepancy. For scenario 1, these
settings produced correlations (spatially, over the land-based pixels)
of the proxy and the focal process that varied (over the 10 replications)
between 0.47 and 0.87 (median 0.67) (with a range of correlations
of 0.76-0.88 when excluding the large-scale discrepancy), somewhat
higher than correlations of 0.58 and 0.52 observed for the real proxies.
In the replicates, larger correlations occurred when the residual
spatial component of the true process and the large-scale discrepancy
were more positively spatially correlated in a given realization,
which occurs by chance because both vary at large spatial scale. Prior
distributions were the same as described in the Appendix for the real
analyses.

\subsection{Results}

I summarize the results based on mean square prediction error (MSPE)
for the land-based pixels, averaged over the 10 replicates (Table
\ref{table:mspe}). Note that it is difficult to compare MSPE values
for different scenario-model combinations in the table because the
variability across replicates is large compared to the differences
between some of the combinations. To address this, in the text below,
I compute standard errors of mean MSPE differences between scenario-model
combinations based on paired replicates.

In general, the model is able to predict spatial variation in the
true process reasonably well in these settings, but it does not exploit
the information in the proxy effectively. Comparing the full model
to a model that does not include the proxy, there is no improvement
in MSPE. In scenarios 1, 3, 4, and 5, the mean difference (standard
error of the difference) over 10 replications for full model minus
reduced model is 0.02 (0.04), 0.10 (0.05), 0.09 (0.05), and 0.02 (0.04).
Note that even for scenario 3, in which the proxy has no spatial discrepancy,
the mean MSPE is larger for the full model. This indicates that lack
of identifiability plays an important role here, with the full model
having difficulty decomposing the spatial variation into signal and
noise. Covariate coefficient estimates are quite sensitive to the
inclusion of the proxy, indicating that there is substantial trading
off of variability between the discrepancy, residual spatial, and
covariate components of the model. In the full model, the posterior
mean of $\beta_{1}$ generally ranges between 0.4 and 0.7, an attenuation
relative to the true value of one, with the model attributing some
of the signal in the proxy to the discrepancy term. In addition to
these general identifiability issues, I note a more subtle difficulty
in making use of the proxy to resolve small-scale spatial features
(such as those induced in the simulations by the covariates). Small-scale
variation not explained by the measured covariates can only be resolved
as part of the pollution process through the residual spatial term,
$g(\cdot)$. However, the spatial process prior, reflecting the natural
Bayesian complexity penalty, penalizes complexity in $g(\cdot)$.
Instead, the simulations indicate that the discrepancy term picks
up the small-scale proxy signal, as the discrepancy term already needs
to operate at a small scale to capture the small-scale proxy discrepancy. 

\begin{table}

\caption{Average MSPE (standard error) for a variety of model structures for
the six simulation scenarios.}
\label{table:mspe}

\begin{tabular}{|>{\centering}p{1.5in}|c|c|c|c|c|c|}
\hline 
 & \multicolumn{6}{c|}{{\small Simulation scenario}}\tabularnewline
\hline 
{\small Model structure} & {\small 1} & {\small 2} & {\small 3} & {\small 4} & {\small 5} & {\small 6}\tabularnewline
\hline 
\hline 
{\small Full model} & {\small 1.09 (0.09)} & {\small 1.03 (0.07)} & {\small 1.17 (0.09)} & {\small 1.16 (0.09)} & {\small 1.09 (0.09)} & {\small 1.67 (0.13)}\tabularnewline
\hline 
{\small No proxy data (Scenarios 1-5 are the same)} & \multicolumn{5}{c|}{{\small 1.07 (0.08)}} & {\small 1.51 (0.11)}\tabularnewline
\hline 
{\small No discrepancy term} & {\small 2.05 (0.11)} & {\small 3.17 (0.17)} & {\small 0.52 (0.05)} & {\small 1.87 (0.12)} & {\small 1.38 (0.08)} & {\small not run$^{1}$}\tabularnewline
\hline 
{\small Discrepancy assumed large scale} & {\small 1.83 (0.19)} & {\small 2.36 (0.21)} & {\small 0.84 (0.08)} & {\small 0.88 (0.07)} & {\small 1.80 (0.17)} & {\small not run$^{1}$}\tabularnewline
\hline 
{\small Fixing $\beta_{1}=1$} & {\small 0.90 (0.06)} & {\small 1.43 (0.09)} & {\small 0.88 (0.05)} & {\small 0.88 (0.05)} & {\small 0.90 (0.06)} & {\small not run$^{1}$}\tabularnewline
\hline 
{\small Proxy used as covariate} & {\small 1.06 (0.09)} & {\small 1.09 (0.08)} & {\small 0.86 (0.05)} & {\small 0.96 (0.05)} & {\small 1.04 (0.07)} & {\small 1.57 (0.11)}\tabularnewline
\hline 
\end{tabular}

$^{1}$These scenario-model combinations were not run as as they were
expected to add little information to the simulation study.
\end{table}

Adding constraints to the model can improve the situation in some
cases (Table \ref{table:mspe}). When there is no small-scale discrepancy,
forcing $\phi(\cdot)$ to vary only at a large scale (by fixing $\kappa$
to be 1000) reduces MSPE relative to not constraining $\phi(\cdot)$
(mean differences of -0.33 (0.08) and -0.29 (0.12) for scenarios 3
and 4, respectively). Fixing $\beta_{1}=1$ also generally improves
MSPE relative to not constraining $\beta_{1}$ (mean differences of
-0.19 (0.08), -0.29 (0.09), -0.28 (0.09), and -0.19 (0.08) for scenarios
1, 3, 4, and 5, respectively). And not surprisingly, MSPE is lowest
(0.52) in scenario 3 (no spatial discrepancy) when the discrepancy
term is excluded. Unfortunately, in real applications, one does not
know when the addition of constraints to improve identifiability is
reasonable (although cross-validation can help to assess this). For
example, when there is small-scale discrepancy, forcing $\phi(\cdot)$
to vary only at a large scale greatly increases MSPE relative to not
constraining $\phi(\cdot)$ (mean differences of 0.75 (0.24), 1.33
(0.23), and 0.71 (0.23), for scenarios 1, 2, and 5 respectively),
as does removing $\phi(\cdot)$ entirely (mean differences of 0.96
(0.16), 2.14 (0.20), 0.29 (0.11) for scenarios 1, 2, and 5, respectively). 

Including the proxy as a covariate does not substantially reduce MSPE
relative to not including the proxy when there is small-scale discrepancy
(mean differences of -0.01 (0.03) and -0.03 (0.12) for scenarios 1
and 5, respectively) (Table \ref{table:mspe}). However, when there
is no discrepancy or only large-scale discrepancy, then including
the proxy as a covariate does reduce MSPE (mean differences of -0.22
(0.09) and -0.11 (.10) for scenarios 3 and 4, respectively, but note
the large standard error for 4), with the model able to account for
large-scale discrepancy in scenario 4 through $g(\cdot)$. The model
can account for a large proportion of the spatial variation without
the proxy, and the proxy is correlated with the measured covariates.
Therefore, it is not surprising that using the proxy, which is essentially
a measurement-error-contaminated variable, as a covariate is not always
helpful.

In scenario 6, with the smaller sample size, the full model and the
model using the proxy as a covariate are still unable to improve upon
the reduced model without the proxy (mean differences of 0.15 (0.16)
and 0.06 (0.08) relative to the model without the proxy) (Table \ref{table:mspe}).
This suggests that even with sparse observations, discrepancy poses
a challenge to taking advantage of the proxy when the reduced model
has reasonable predictive ability ($R^{2}$ values ranging between
0.16 and 0.85 with a median of 0.68). 

Note that the difficulty in exploiting the proxy occurs with correlations
between the proxy and the truth that are on average higher than those
seen in the real analyses of Section \ref{sec:Analysis-of-particulate}.
This suggests that one needs a high-quality proxy when a model without
the proxy performs reasonably well on its own. In the presence of
discrepancy, the model discounts the proxy (recall the attentuated
estimates of $\beta_{1}$), but error in the proxy still 'leaks' into
the focal process inference. Whether this leakage outweighs any benefit
of proxy signal for prediction depends in a complicated fashion on
the amount of information available in the observations and the signal
to noise ratio in the proxy. On a positive note, the simulations indicate
that the model is able to screen out proxies that are not sufficiently
informative, with little decrease in predictive performance, despite
the imbalance between gold standard and proxy sample sizes. On a negative
note, in the simulations neither the discrepancy nor $\beta_{1}$
vary with the covariates, which is not realistic in the air pollution
context and would likely make it harder to take advantage of a proxy
by introducing non-identifiability between the covariates and the
discrepancy. 

Finally, I briefly consider prediction uncertainty (for scenario 1).
The full model and the model including the proxy as a covariate have
somewhat smaller prediction standard deviations (averaged over the
land-based pixels) than the model without the proxy (0.89 and 1.01
compared to 1.04), but coverage is also lower (0.86 and 0.90 compared
to 0.91).

\begin{figure}
\includegraphics[scale=0.75]{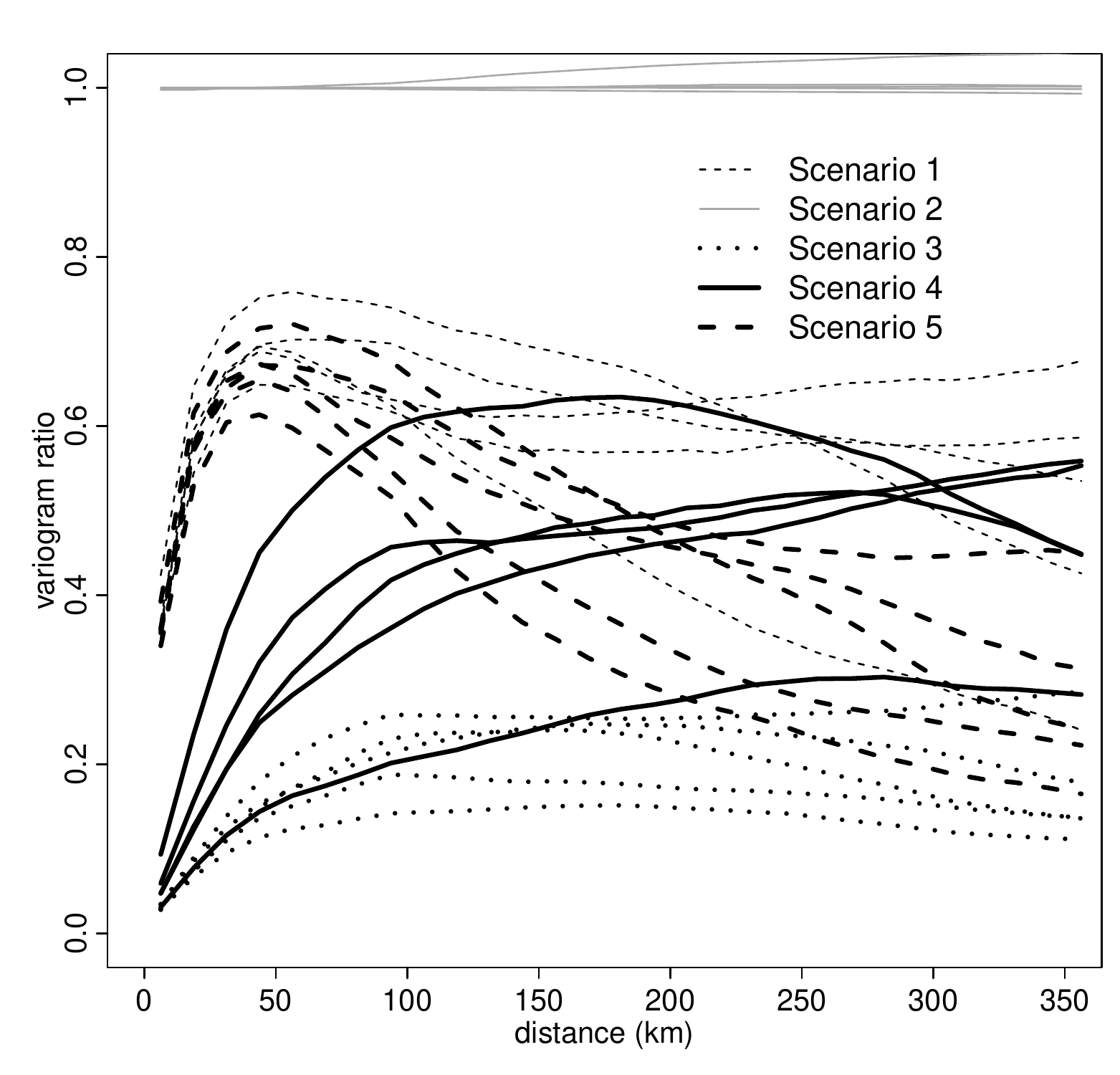}

\caption{Discrepancy scale diagnostic for the model fit to simulated data (the
first five of the 10 replications for clarity) under various scenarios:
(1) large- plus small-scale discrepancy (thin dashed line), (2) uninformative
proxy (thin grey line), (3) no spatial discrepancy (dotted line),
(4) large-scale discrepancy only (thick line), and (5) small-scale
discrepancy only. \label{sims-stein}}
\end{figure}

Fig. \ref{sims-stein} shows the discrepancy scale diagnostic (Section
\ref{sub:Spatial-discrepancy-diagnostic}) for the full model under
each scenario. In general, there appears to be reasonable, but imperfect,
information in the diagnostic, with the expected relationship with
scale given the data generation for a given scenario. That is, when
there is more discrepancy at large than small scales the ratio should
increase with distance (scenario 4) and when there is more discrepancy
at small than large scales, it should decrease (scenario 5). Unfortunately,
even when there is no true discrepancy at a scale the diagnostic estimates
discrepancy, which is caused by the identifiability problems. This
suggests that one should treat the diagnostic as indicative, rather
than conclusive, about the scales of discrepancy. Also, the diagnostic
results suggest caution in interpreting the diagnostic at the shortest
distances, given the drop in the diagnostic even for the scenarios
with small scale discrepancy (scenarios 1 and 5).

\section{Analysis of particulate matter\label{sec:Analysis-of-particulate}}

\subsection{Spatial analysis of MODIS AOD in the mid-Atlantic region\label{sub:aodResults}}

I fit separate spatial models for each month of 2004 for the mid-Atlantic
region with AOD from the MODIS instrument as the proxy. I used a regular
four km grid of dimension 175 by 100 as the resolution of $L(\cdot)$,
$g(\cdot)$, and $\phi(\cdot)$. MODIS AOD is misaligned with respect
to this grid and for different satellite orbits on different days,
the pixels shift spatially. Therefore, I considered the overlap of
all the pixels in an orbit with the four km grid, assigning to each
grid cell, $\bld{s_{j}}$, the value of the MODIS pixel in which the
cell centroid falls. Taking the retrievals assigned to each cell,
I then averaged to the monthly level for each cell. More sophisticated
approaches are possible \citep{Mugg:etal:2000}, but for these purposes,
this ad hoc realignment retained the essential character of the AOD
retrievals and reduced computations. I attempted to account in part
for informatively-missing AOD due to cloud cover \citep{Paci:etal:2008}
by including a smooth regression term, $\beta_{a}(x_{a})$, in the
additive mean for the proxy, thereby making a missing-at-random assumption.
$x_{a}$ is the average cloud cover over the month for each location,
based on the cloud screen variable from the Geostationary Operational
Environmental Satellite. $g(\cdot)$ was modeled using the thin plate
spline mixed model representation. Distances to nearby roads in two
road size classes, calculated for each monitor location, were used
as smooth spline terms, $\beta_{y,p}(x_{y,p}),\, p=1,2$, while elevation,
population density, road density in three road size classes, and area
emissions were estimated at the four km resolution using a Geographic
Information System and used as smooth spline terms, $\beta_{L,p}(x_{L,p}),\, p=1,\ldots6$.
Distance to point source emissions (those greater than five tons per
year) was handled as a smooth function of distance, adding over the
sources within 100 km of the observation (or grid cell), following
the methodology of \citet[Appendix D]{Paci:Liu:2011}. 

I considered both raw MODIS AOD and a 'calibrated' MODIS AOD that
adjusts off-line for the effects of boundary layer height and relative
humidity, as well as large-scale temporal and spatial adjustments,
based on comparison of co-located daily PM and AOD values \citep{Paci:Liu:2009}.
Results were similar and are presented here for raw MODIS AOD.

\begin{figure}
\includegraphics[scale=0.8]{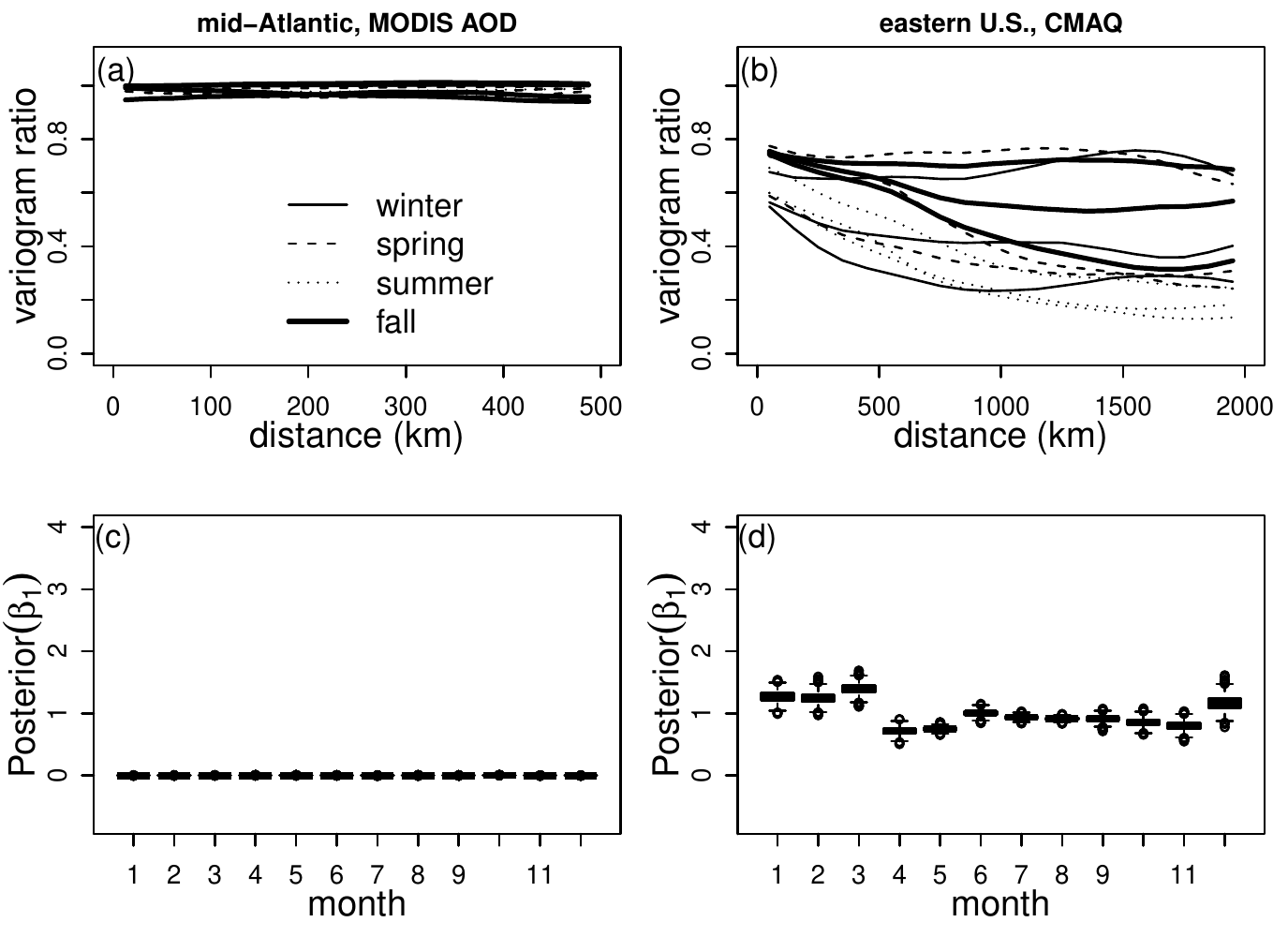}

\caption{Discrepancy scale diagnostic plots (top row) and boxplots of posterior
estimates of $\beta_{1}$ (bottom row) by month for analysis of MODIS
AOD as the proxy in 2004 in the mid-Atlantic (left panel) and CMAQ
as the proxy in 2001 in the eastern U.S. (right panels).\label{vgBeta1}}
\end{figure}

\begin{figure}
\includegraphics[scale=0.52]{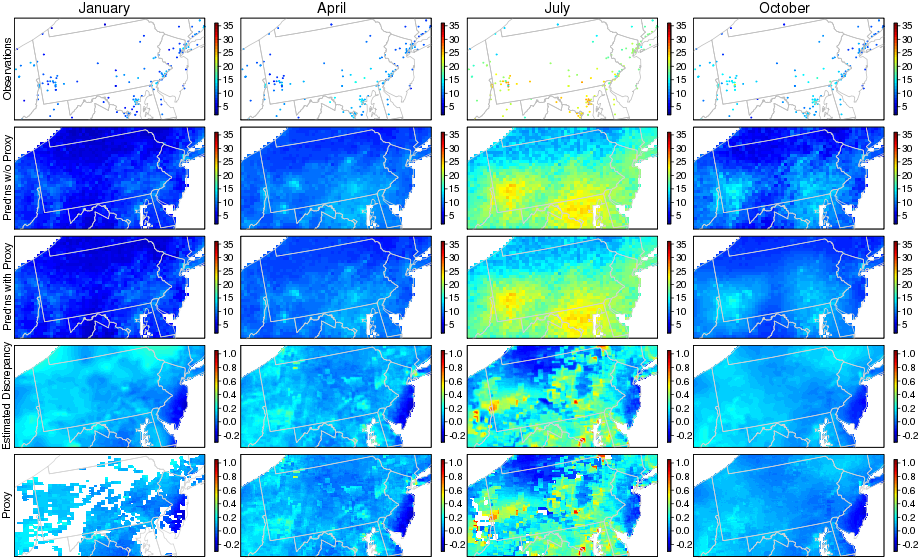}

\caption{For the mid-Atlantic region of the U.S., PM observations (first row),
model PM predictions excluding MODIS AOD (second row), model predictions
including AOD (third row), estimated discrepancy in the model with
AOD (fourth row), and AOD values (fifth row) for four months in 2004:
January, April, July, and October (columns). White space in the land
areas in the fifth row indicate AOD values missing because of cloud
cover for all AOD retrievals in the month.\label{fig:predPlots1}}
\end{figure}

The estimated values of the regression coefficient, $\beta_{1}$,
relating the proxy to the focal process were essentially zero in every
month (Fig. \ref{vgBeta1}c), indicating that the model found no relationship
between the proxy and the PM process, ignoring the proxy in predicting
PM. The spatial discrepancy plot indicates that the proxy is explained
by the discrepancy term rather than the PM process at all scales (Fig.
\ref{vgBeta1}a). These results are consistent with those found in
\citet{Paci:Liu:2009}. Fig. \ref{fig:predPlots1} shows predictions
from the model when excluding and including AOD, with very similar
predictions. Correspondingly, (ten-fold) cross-validative predictive
assessment indicated that including the proxy did not improve predictive
performance (Table \ref{tab:CVR2}). (Yearly prediction appeared slightly
worse when excluding the proxy, but this is likely within the uncertainty
of the predictive assessment, which is difficult to quantify given
the correlation structure of the data). Results were qualitatively
similar when considering only held-out monitors in more rural areas,
suggesting that AOD is not adding information in areas far from monitors
(Table \ref{tab:CVR2}). In light of the simulation results, these
results suggest that we are in the regime where the proxy is not sufficiently
informative about the process of interest (relative to the strength
of information in the gold standard data) to be able to exploit the
information in the proxy indicated by the AOD-PM correlation of 0.58.

\begin{table} \caption{Cross-validation predictive ability, $R^{2}$ (RMSPE), for monthly and yearly (average of the 12 monthly values) predictions, with and without proxy information, over the spatio-temporal domains of the analyses. \label{tab:CVR2}} \begin{tabular}{|p{1.5in}|p{1.1in}|p{1.3in}|p{1.3in}|} \hline  Time scale & Proxy inclusion & mid-Atlantic, 2004, MODIS AOD & eastern U.S., 2001, CMAQ\tabularnewline \hline \hline  \multicolumn{4}{c}{All monitors} \tabularnewline \hline Monthly prediction$^{1}$ & With proxy & 0.806 (1.80)  & 0.827 (1.71)\tabularnewline \hline & Without proxy & 0.808 (1.79)  & 0.826 (1.72)\tabularnewline \hline Yearly prediction$^{2}$ & With proxy & 0.667 (1.00)$^{3}$ & 0.800 (1.21)\tabularnewline \hline & Without proxy & 0.650 (1.03)$^{3}$ & 0.835 (1.09)\tabularnewline \hline  \multicolumn{4}{c}{Monitors generally isolated from other monitors$^4$}  \tabularnewline \hline Monthly prediction$^{1}$ & With proxy & 0.830 (1.73)  & 0.739 (2.12)\tabularnewline \hline & Without proxy & 0.830 (1.72)  & 0.781 (1.94)\tabularnewline \hline Yearly prediction$^{2}$ & With proxy & 0.669 (1.17) &  0.710 (1.58)\tabularnewline \hline & Without proxy & 0.624 (1.25)  & 0.826 (1.22)\tabularnewline \hline \end{tabular} \\ \renewcommand{\baselinestretch}{1.0} \small $^{1}$ Including monthly averages based on at least five daily observations. $^{2}$Including yearly averages (averages of monthly values) based on at least nine months with at least five daily observations. $^{3}$ Excludes one site outside Pittsburgh just downwind of a major industrial facility. $^4$ Monitors in four km grid cells with fewer than 187.5 people per square km=3000 people per cell.  \renewcommand{\baselinestretch}{1.5}
 \end{table}

\subsection{Spatial analysis of CMAQ output in the eastern United States\label{sub:subnatl}}

I fit separate monthly spatial models for 2001 for the entire eastern
U.S. with CMAQ output as the proxy. For this analysis, I extended
the four km grid over the the eastern U.S. (roughly the area east
of $100^{\degr}$W longitude), giving a grid of dimension 669 by 677.
I used the 36 km CMAQ grid of dimension 73 by 77 for $\phi(\cdot)$.
For this larger domain with more complexity in the pollution surface,
I represented $g(\cdot)$ on this same 36 km grid using the TPS-MRF
representation rather than the mixed model thin plate spline. The
latter would have required a computationally burdensome increase in
the number of basis coefficients. Note that I relied on the same covariates
used in the mid-Atlantic analysis to represent small-scale residual
spatial variability, such that representing $g(\cdot)$ on the 36-km
CMAQ grid is sufficient, based on posterior assessment of the scale
of $g(\cdot)$. To calculate the PM likelihood I used the covariate
values for the four km grid cell that each observation falls in, while
using the value of $\bld{g}$ from the 36 km grid cell the observation
falls in. For the CMAQ likelihood, the CMAQ value for a given 36 km
grid cell was related to the covariate effects on the four km grid
by weighted averaging, with weights based on the overlap between a
CMAQ grid cell and the four km grid cells. 

$\beta_{1}$ was estimated to be large (between three and seven),
with $\bld{\phi}$ strongly negatively correlated with $\bld{L}$
and with $\bld{\phi}$ having very large negative values to offset
the large positive values of $\beta_{1}\bld{L}$. This appeared to
be driven by CMAQ overpredicting in some urban areas, with CMAQ estimating
the urban to rural gradient as being much stronger than apparent in
the observations. $\bld{\phi}$ then adjusts for the impact outside
urban areas of large values of $\beta_{1}$, highlighting the lack
of identifiability between the discrepancy and the PM process. To
address this in an ad hoc manner and identify $\bld{\phi}$ as the
orthogonal variation in the proxy not accounted for in the PM process,
I carried out an ad hoc orthogonalization of $\bld{\phi}$ and $\bld{L}$
within each step of the MCMC. While a formal orthogonality constraint
on $\bld{\phi}$ and $\bld{L}$ is technically appealing, this simple
approach was effective in practice, and predictive results were similar
with and without the orthogonalization. 

\begin{figure}
\includegraphics[scale=0.52]{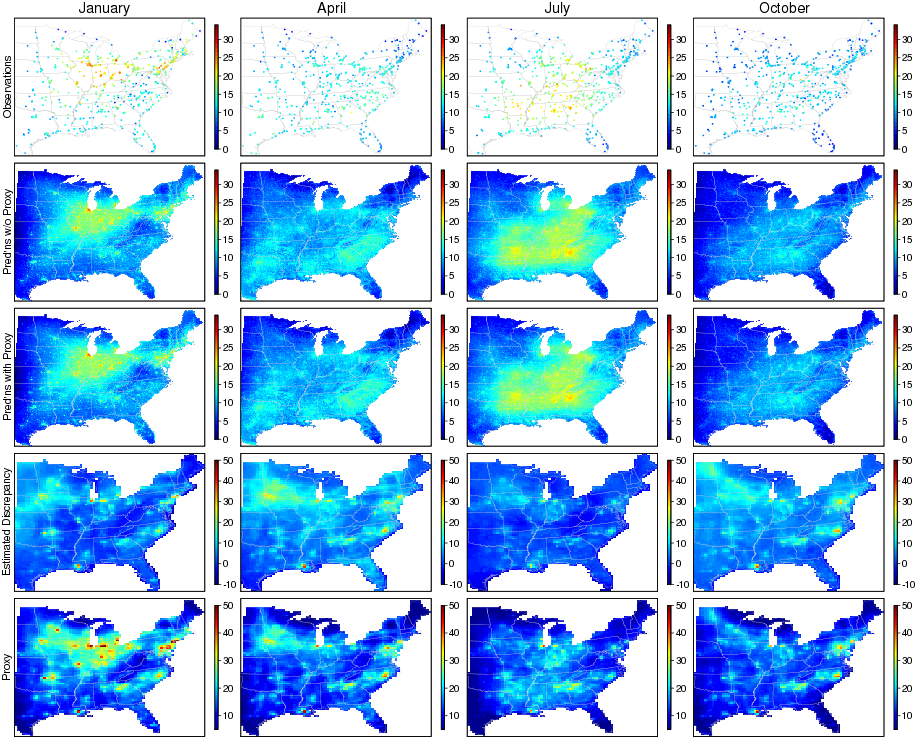}

\caption{PM observations (first row), model PM predictions, excluding CMAQ
(second row), model predictions including CMAQ (third row), estimated
CMAQ discrepancy (fourth row), and CMAQ values (fifth row) for four
months in 2001: January, April, July, and October (columns). Note
that the scale for the CMAQ values is different than for the PM observations.
Also note that in January, some CMAQ values larger than 50 $\mu g/m^{3}$
(up to 75 $\mu g/m^{3}$) are truncated to 50 $\mu g/m^{3}$. \label{fig:predPlots3}}
\end{figure}

Using this orthogonalized specification, the estimated values of $\beta_{1}$
were generally near one, as one would hope (Fig. \ref{vgBeta1}d).
Somewhat more of the variability in CMAQ at larger scales was associated
with the PM process than variability at smaller scales, suggesting
that CMAQ is better able to resolve regional variability than more
local variability (Fig. \ref{vgBeta1}b). This is consistent with
the discrepancy surfaces shown in Fig. \ref{fig:predPlots3}, which
include hotspots at the scale of states or portions of states that
show no evidence of being real hotspots based on the observations,
such as over Iowa/southern Minnesota and eastern North Carolina. Use
of CMAQ as a proxy did not improve predictive ability (Table \ref{tab:CVR2}). 

Considered in light of the simulation results, these results suggest
that we are again in the regime where the proxy is not sufficiently
informative about the process of interest (relative to the strength
of information in the gold standard data) to be able to exploit the
information in the proxy indicated by the CMAQ-PM correlation of 0.52.
Given the apparent impact of extreme CMAQ predictions, further model
refinement, such as letting $\beta_{1}$ vary with covariates such
as population density, might improve our ability to extract information
from the proxy.

As a final assessment, I included CMAQ as a simple regression term,
finding the cross-validation $R^{2}$ (RMSPE) to be 0.849 (1.60) for
monthly prediction and 0.849 (1.05) for yearly prediction, both slightly
better than modeling without the proxy (Table \ref{tab:CVR2}) but
suggesting that there is limited additional information given the
other terms in the model. Similarly modest improvements were seen
when restricting to more rural sites. The posterior mean regression
coefficients for CMAQ-estimated PM were between 0.48 and 0.89 for
the 12 months, with the average of those 12 posterior means being
0.67.

I also considered whether CMAQ might be more helpful in a setting
with sparse observations by artificially using only 100 monitors as
the PM dataset, but found that prediction for the remaining monitor
locations was very poor, consistent with the simulation results, because
the model was not able to adjust for the CMAQ discrepancy as well
as with more dense data.

\section{Discussion}

The model developed here is a spatial latent variable model, in which
the proxy is decomposed into signal for the process of interest and
noise, which raises the obvious concern of identifiability. Sufficient
gold standard observations are required for the implicit calibration
of the proxy that is done within the model fitting. The penalized
model identifies the discrepancy and the focal process based on a
tradeoff between goodness of fit for the observations and for the
proxy values and penalization of the latent spatial processes. This
tradeoff is likely quite sensitive to model specification (note the
trading off that occurred in the CMAQ analysis) and to the relative
richness of observation and proxy data. The limited identifiability
of the spatial processes in the model helps to explain the difficulty
in exploiting the proxies seen in the simulations and the real analyses.
Large-scale discrepancy occurring at a similar scale as the process
of interest can leak into the estimated process. This stands in contrast
to regression models in which fitting involves a projection that fully
filters out the effect of covariates included for confounding control.
When there is both discrepancy and signal in the proxy at smaller
scales, the model has no way to distinguish these based on sparse
observations. The discrepancy necessarily distorts the predictions.
One can interpret the regression coefficient that relates the focal
process and the proxy as a shrinkage parameter that weights the impact
of the proxy on the focal process in the context of a complicated
bias-variance tradeoff. Whether the proxy can improve prediction depends
on the relative strength of the signal and noise in the proxy and
the strength of information in the observations.

Despite these concerns about identifiability and sensitivity to model
structure, the decomposition of signal and noise is a fundamental
task when using proxies. One doesn't know the quality of the proxy
or the structure of any discrepancy and must assess this, ideally
as a function of scale. As in the computer models literature \citep{Baya:etal:2007},
the lack of identifiability in the modeling follows from the scientific
context. In contrast, a priori constraints on the flexibility of the
discrepancy process incorporated into some data fusion approaches
make strong assumptions about proxy quality. The model presented here
attempts to assess proxy quality and to make use of available information
in the proxy without prejudging that spatial correlation in the proxy
reflects spatial signal for the process of interest. The results highlight
the difficulty in actually improving predictions when including proxy
information for which discrepancy is a concern and the need for new
approaches that do not make strong assumptions about proxy quality
but are better able to exploit information in the proxy. 

The model accounts for errors and local variability in the observations,
as well as spatial and temporal misalignment between observations
and proxy. This should help to avoid concluding that a proxy is not
helpful because of noise or fine-scale variability in the 'gold standard',
but I cannot completely rule out this possibility. However, I note
that comparisons of daily maps of proxy values and observations indicate
a great deal of mismatch between the two, of particular note on days
when the observations show large-scale variation but little local
variation. On such days the patterns in true PM are well-identified
from the observations, yet the proxies often did a poor job of capturing
that variation. It is possible that the proxy variables might add
useful information in the absence of the GIS-based covariates, but
part of the purpose of the covariates is to account for local variation
in the observations that contributes to discrepancy between proxy
and observations. Local variation that is not accounted for in the
model structure is likely to mask any real relationship between the
proxy and the true process of interest at the grid scale and may lead
to discounting real information in the proxy. 

As noted by a referee, it may seem curious to use a MRF representation
in a context in which spatial scale is critical, given the lack of
a spatial range parameter in the MRF. A recently-developed alternative
to the TPS-MRF is the MRF approximation to GPs in the \matern class
\citep{Lind:etal:2011}. However, even GP approaches only explicitly
model variation at a single scale. (Note they will also follow larger-scale
variation except in data gaps.) To what extent might the limitations
of the TPS-MRF prevent us from better exploiting the proxy? When there
is large-scale discrepancy or no discrepancy, there is no reason to
think that a GP or multiresolution model for the discrepancy would
do any better in handling the identifiability problem. However, when
only small-scale discrepancy is present, a multiresolution approach
that imposes identifiability based on differentiating spatial scales
might do a better job of preventing large-scale signal from being
represented in the discrepancy term. More exploration of multiresolution
approaches would be worthwhile, but they seem unlikely to deal substantially
better with identifiability issues when the signal and noise vary
at similar scales.

Rather than treating proxy information as data, \citet{Berr:etal:2010}
proposed instead to regress on the proxy, which also greatly reduces
computational complexity. In the simulations, I found that regressing
on the proxy in some cases (no discrepancy or large-scale discrepancy
only) improved predictive ability and in others (when small-scale
discrepancy was present) had little effect. (Note that the presence
of covariates in my settings likely decreases the potential for improvements
from including the proxy, in comparison with the improved prediction
seen in \citet{Berr:etal:2010} in a scenario without covariates.)
So if one is purely interested in prediction and not in understanding
the structure of the proxy, a simple regression approach may be best.
An important drawback is that for proxies such as remote sensing retrievals
affected by cloud cover, missing observations are a problem. This
then requires an imputation of some sort, bringing one back to modeling
the proxy. 

The second potential drawback is more subtle and involves difficulties
in decomposing the proxy into scales when the relationship with the
process of interest varies by scale. If the proxy captures small-scale
patterns well, but misses large-scale patterns, then the additive
spatial process term used in \citet{Berr:etal:2010} can adjust for
this discrepancy, leveraging the proxy to improve prediction of small-scale
variation, a phenomenon seen also in my simulations. Concern arises
when the proxy captures large-scale patterns well, but the small-scale
patterns in the proxy are predominantly noise, a plausible scenario
in applications. In such a setting, the model may estimate a large
regression coefficient for the proxy because of the large-scale association.
The resulting prediction will be reasonable at the large scale but
will also include all the small-scale spatial discrepancy in the proxy,
which will be interpreted by the analyst as signal, with the proxy
assumed to be of high quality at small scales. The additive spatial
process in the model will not be able to correct for this small-scale
discrepancy because of the sparsity of the gold standard observations
relative to the proxy. Alternatively, the small-scale discrepancy
in the proxy may act as measurement error, causing attenuation in
the regression coefficient and limiting any information gain from
the large-scale signal in the proxy. One would like to carry out a
decomposition of the scales of variability in the proxy, but the regression
approach conditions on all the scales of the proxy. 

From an interpretation standpoint, the approach proposed in this work
cleanly distinguishes between discrepancy and signal in principle,
albeit not in practice. A simpler alternative would be to explicitly
decompose the proxy into two or more scales and include each component
as a separate regression term, allowing the model to learn which scales
are correlated with the observations. One might successively smooth
the proxy with spatial smooth terms with fixed degrees of freedom
or use kernel average predictors, as suggested in \citet{Heat:Gelf:2011}.

\section*{Acknowledgements}

I thank Yang Liu for processing the MODIS AOD and CMAQ output, as
well as for collaboration on the overall project that led to this
work, and Steve Melly for GIS processing. I thank Atmospheric and
Environmental Research and the Electric Power Research Institute for
providing the CMAQ output.  This work was supported by grant 4746-RFA05-2/06-7
from the Health Effects Institute

\bibliographystyle{/home/paciorek/scf/latex/RSSstylefile/Chicago}

\bibliography{/home/paciorek/scf/bibfiles/abbrev.stat,/home/paciorek/scf/bibfiles/abbrev.other,/home/paciorek/scf/bibfiles/spatfit,/home/paciorek/scf/bibfiles/spatstat,/home/paciorek/scf/bibfiles/statgeneral,/home/paciorek/scf/bibfiles/paciorek,/home/paciorek/scf/bibfiles/ml,/home/paciorek/scf/bibfiles/thesis,/home/paciorek/scf/bibfiles/env}

\section*{Appendix: Model structures and fitting details}

\subsection*{Spatial model structure for the mid-Atlantic analysis}

The basic model is a model with two likelihoods and additive mean
terms, in particular 

\begin{eqnarray*}
\bld{Y} & \sim & \mathcal{N}_{n_{Y}}(\bld{Z}_{y}\bld{b}_{y}+\bld{Z}_{L}\bld{b}_{L}+\bld{P}_{\delta}\bld{\delta},\bld{V}_{Y})\\
\bld{A} & \sim & \mathcal{N}_{n_{A}}(\bld{P}_{\phi}\bld{\phi}+\bld{Z}_{a}\bld{b}_{a}+\beta_{1}\bld{P}_{A}\bld{Z}_{L}\bld{b}_{L},\bld{V}_{A})\\
\bld{\delta} & \sim & \mathcal{N}(\bld{0},\sigma_{h}^{2}\bld{I})\\
\bld{\phi} & \sim & \mathcal{N}_{m-3}(\bld{0},(\kappa\bld{Q})^{-})\\
\bld{b}=\{\bld{b}_{y},\bld{b}_{L},\bld{b}_{a}\} & \sim & \mathcal{N}(\bld{0},\bld{\Lambda})
\end{eqnarray*}
where $\bld{Z}_{y}\bld{b}_{y}$ is the matrix representation of $\sum_{p}\beta_{y,p}(x_{y,p})$
and $\bld{Z}_{L}\bld{b}_{L}$ is the matrix representation of $\sum_{p}\beta_{L,p}(x_{L,p})+g(s)$
with $\bld{b}_{L}$ the collection of combined coefficients for the
regression smooths and the spatial term, as well as including an intercept
for $\bld{Y}$. Similarly, $\bld{Z}_{a}\bld{b}_{a}$ represents the
influence of explanatory variables for the proxy unrelated to latent
PM (cloud cover in the case of the AOD model). $\bld{\delta}$ are
site-specific effects that account for correlation between monitors
placed at the same site. I denote the variance of $\bld{\phi}$ using
the generalized inverse to indicate that the prior is proper in an
$m-3$ dimensional space based on the construction of the TPS-MRF,
fixing the mean and coefficients for linear terms of the spatial coordinates
to zero. In the sampling, the three parameters are identified by the
likelihood for $\bld{A}$, so I sample these parameters implicitly
as part of $\bld{\phi}$ and therefore omit a separate intercept for
$\bld{A}$. Given the limited number of observation locations, I use
five knots for each regression smooth and 55 knots for the spatial
residual. Knots were placed either uniformly over the range of covariate
values or at equally-spaced quantiles to achieve reasonable spread
over the covariate spaces, but given the use of penalized splines,
results should be robust to the exact placement of knots. $\bld{\Lambda}$
is the prior covariance matrix of $\bld{b}$ (non-informative for
the fixed effect components and with exchangeable priors amongst the
coefficients for a given regression smooth term, following \citet{Rupp:etal:2003}).

I integrate over the conditional normal distribution of $\bld{\phi}$,
with mean $\bld{M}_{\phi}=\bld{V}_{\phi}(\bld{P}_{A}^{T}\bld{V}_{A}^{-1}(\bld{A}-\bld{Z}_{a}\bld{b}_{a}-\bld{P}_{A}\bld{Z}_{L}\bld{b}_{L}))$
and variance $\bld{V}_{\phi}=(\bld{P}_{\phi}^{T}\bld{V}_{A}^{-1}\bld{P}_{\phi}+\kappa\bld{Q})^{-1}$
to obtain
\[
\bld{A}\sim\mathcal{N}_{n_{A}-3}(\bld{Z}_{a}\bld{b}_{a}+\beta_{1}\bld{P}_{A}\bld{Z}_{L}\bld{b}_{L},\bld{\Sigma}_{A}).
\]
Here $\bld{\Sigma}_{A}^{-1}=\bld{V}_{A}^{-1}-\bld{V}_{A}^{-1}\bld{P}_{\phi}\bld{V}_{\phi}\bld{P}_{\phi}^{T}\bld{V}_{A}^{-1}$
and $|\bld{\Sigma}_{A}|^{-\frac{1}{2}}$ can be expressed as 
\[
\frac{1}{|\bld{\Sigma}_{A}|^{\frac{1}{2}}}=\frac{\kappa^{\frac{m-3}{2}}|\bld{Q}|^{\frac{1}{2}}}{|\bld{V}_{A}|^{\frac{1}{2}}|\bld{V}_{\phi}^{-1}|^{\frac{1}{2}}}.
\]
Note that the impropriety in the prior for $\bld{\phi}$ carries over
into this marginal likelihood for $\bld{A}$, resulting in $m-3$
rather than $m$ in the exponent of $\kappa$ and in $\bld{\Sigma}_{A}^{-1}$
being singular, with three zero eigenvalues, but the subsequent calculations
all involve $\bld{\Sigma}_{A}^{-1}$ so no inversion is needed. Equivalently,
I do not have a legitimate data-generating model for $\bld{A}$ because
of the prior impropriety. Information from three of the linear combinations
in the quadratic form of the marginal likelihood contribute zero to
the marginal likelihood because the variance for those combinations
is infinite. I can avoid calculating the non-existent determinant
of $\bld{Q}$ because this is a constant with respect to the model
parameters. Note that the impropriety is analogous to that in the
marginal likelihood obtained from integrating over the mean in a simple
normal mean problem with an improper prior for the mean. 

I then integrate over the joint distribution for $\bld{b}$, where
I construct $\bld{Z}_{Y}$ and $\bld{Z}_{A}$ such that $\bld{Z}_{Y}\bld{b}=\bld{Z}_{y}\bld{b}_{y}+\bld{Z}_{L}\bld{b}_{L}$
and $\bld{Z}_{A}\bld{b}=\bld{Z}_{a}\bld{b}_{a}+\bld{P}_{A}\bld{Z}_{L}\bld{b}_{L}$
by adding columns with all zeroes as necessary. The conditional distribution
for $\bld{b}$ is normal with mean, $\bld{M}_{b}=\bld{V}_{b}(\bld{Z}_{Y}^{T}\bld{V}_{Y}(\bld{Y}-\bld{P}_{\delta}\bld{\delta})+\bld{Z}_{A}^{T}\bld{\Sigma}_{A}^{-1}\bld{A})$
and $\bld{V}_{b}=(\bld{Z}_{Y}^{T}\bld{V}_{Y}^{-1}\bld{Z}_{Y}+\bld{Z}_{A}^{T}\bld{\Sigma}_{A}^{-1}\bld{Z}_{A}+\bld{\Lambda}^{-1})^{-1}$.
I collect the remaining parameters as $\bld{\theta}=\{\beta_{1},\sigma_{\smttt{sub}}^{2},\sigma_{\delta}^{2},\sigma_{\epsilon}^{2},\sigma_{A}^{2},\sigma_{\alpha}^{2},\bld{\sigma}_{b,L}^{2},\bld{\sigma}_{b,a}^{2},\bld{\sigma}_{b,y}^{2},\kappa\}$
where the variance components, $\bld{\sigma}_{b,y}^{2}$, $\bld{\sigma}_{b,L}^{2}$,
and $\bld{\sigma}_{b,a}^{2}$, are vectors with one variance component
for each smooth regression term in a given sum of regression smooths
and $\sigma_{\smttt{sub}}^{2}$, $\sigma_{\delta}^{2}$, $\sigma_{\epsilon}^{2}$,
$\sigma_{A}^{2}$, and $\sigma_{a}^{2}$ are parameters used to construct
$\bld{V}_{Y}$ and $\bld{V}_{A}$, described below. The marginal posterior
for the remaining parameters and $\bld{\delta}$ is
\begin{eqnarray}
P(\bld{\theta},\bld{\delta}|\bld{A},\bld{Y}) & \propto & |\bld{\Lambda}|^{-\frac{1}{2}}|\bld{V}_{Y}|^{-\frac{1}{2}}|\bld{\Sigma}_{A}|^{-\frac{1}{2}}|\bld{V}_{b}|^{\frac{1}{2}}P(\bld{\delta})P(\bld{\theta})\cdot\nonumber \\
 &  & \exp\left(-\frac{1}{2}\left((\bld{Y}-\bld{P}_{\delta}\bld{\delta})^{T}\bld{V}_{Y}^{-1}(\bld{Y}-\bld{P}_{\delta}\bld{\delta})+\bld{A}^{T}\bld{\Sigma}_{A}^{-1}\bld{A}-\bld{M}_{b}^{T}\bld{V}_{b}^{-1}\bld{M}_{b}\right)\right),\label{eq:margPost}
\end{eqnarray}
which I use to sample $\bld{\theta}$ via blocked Metropolis. Depending
on the model, in some cases I use a single block and in other cases
subblocks. I use adaptive MCMC to tune the proposal covariance matrix
throughout the chain \citep{Andr:Thom:2008}. 

The key computational impediments involve the determinant of $\bld{V}_{\phi}^{-1}$,
which can be calculated based on sparse matrix operations since both
of its components are sparse; note that $\bld{P}_{\phi}$ is a simple
sparse mapping matrix assigning elements of $\bld{\phi}$ to the proxy
values. Next $\bld{V}_{b}$ is a dense matrix whose size corresponds
to the number of basis coefficients, which can be computationally
burdensome when I use a large number of knots for $\bld{g}$ or the
total number of knots used for all the regression smooth terms is
large. Finally, I must compute $\bld{\Sigma}_{A}^{-1}\bld{A}$ and
$\bld{\Sigma}_{A}^{-1}\bld{Z}_{A}$, the latter being more burdensome
because $\bld{Z}_{A}$ is a matrix with as many non-zero columns as
there are coefficients in $\{\bld{b}_{a},\bld{b}_{L}\}$. Considering
the representation of $\bld{\Sigma}_{A}^{-1}$, note that I can easily
compute $\bld{V}_{A}^{-1}\bld{Z}_{A}$ and then $\bld{P}_{\phi}^{T}\bld{V}_{A}^{-1}\bld{Z}_{A}$
because $\bld{V}_{A}^{-1}$ is diagonal and $\bld{P}_{\phi}^{T}$
is sparse. Next I use sparse matrix operations to solve the system
of equations $\bld{V}_{\phi}\bld{P}_{A}^{T}\bld{V}_{A}^{-1}\bld{Z}_{A}$
(recall that I calculate $\bld{V}_{\phi}^{-1}$ quickly as the sparse
matrix sum of two sparse matrices). I use the spam package in R for
sparse matrix calculations. 

Given the posterior for the remaining parameters (\ref{eq:margPost}),
I can derive the closed form normal conditional distribution for $\bld{\delta}$,
which has mean and variance,
\begin{eqnarray*}
\bld{M}_{\delta} & = & \bld{V}_{\delta}(\bld{P}_{\delta}^{T}\bld{V}_{Y}^{-1}\bld{Y}-\bld{P}_{\delta}^{T}\bld{V}_{Y}^{-1}\bld{Z}_{Y}\bld{V}_{b}(\bld{Z}_{Y}^{T}\bld{V}_{Y}^{-1}\bld{Y}+\bld{Z}_{A}^{T}\bld{\Sigma}_{A}^{-1}\bld{A}))\\
\bld{V}_{\delta}^{-1} & = & \sigma_{\delta}^{2}\bld{I}+\bld{P}_{\delta}^{T}\bld{V}_{Y}^{-1}\bld{P}_{\delta}-\bld{P}_{\delta}^{T}\bld{V}_{Y}^{-1}\bld{Z}_{Y}\bld{V}_{b}\bld{Z}_{Y}^{T}\bld{V}_{Y}^{-1}\bld{P}_{\delta}.
\end{eqnarray*}
Sampling from this distribution efficiently involves sparse matrix
calculations similar to those just described. 

Posterior samples of $\bld{\phi}$ and $\bld{b}$ can be drawn off-line
from the conditional distributions indicated above; I choose to draw
them every 10 MCMC iterations.

$\bld{V}_{Y}$ is modeled using a diagonal heteroscedastic variance,
$(\bld{V}_{Y})_{ii}=\sigma_{\epsilon}^{2}/n_{i}+k(n_{i})\sigma_{\smttt{sub}}^{2}$
where the first term is the variance of the sum of independent daily
instrument errors. The second reflects subsampling variability in
the average of $n_{i}$ instrument-error-free daily values relative
to the average of the error-free daily values over all the days in
the month, $\mbox{Var}(\sum_{d\in\smttt{subsample}}L_{i,d}/n_{i}-\sum_{d}L_{i,d}/n_{\smttt{month}})$,
under the simplifying assumption of independence between true daily
pollution values, $L_{i,d}\stackrel{iid}{\sim}\mathcal{N}(L(\bld{s}_{i}),\sigma_{\smttt{sub}}^{2})$,
which gives $k(n_{i})=\frac{1}{n_{i}}-\frac{1}{n_{\smttt{month}}}$,
where $n_{\smttt{month}}$ is the number of days in the month. Note
that for simplicity I fixed $\sigma_{\epsilon}^{2}$ at $\widehat{\sigma_{\epsilon}^{2}}=1.5$,
estimated in advance from co-located monitors, to enhance identifiability
and because it has a small contribution to the overall error variance.
For monitors not co-located with another monitor, I integrated over
the prior for the $\delta$ values for those sites, which added a
term, $\sigma_{h}^{2}$, to $(\bld{V}_{Y})_{ii}$. In contrast, I
sampled the values of $\delta$ for sites with co-located monitors
within the primary MCMC to avoid introducing off-diagonal elements
into $\bld{V}_{Y}$ as this would have obviated some of the computational
efficiencies in the calculations outlined above. For models involving
CMAQ, which is available for all days, I take $\bld{V}_{A}=\sigma_{A}^{2}\bld{I}$.
For models involving AOD, I use a diagonal heteroscedastic variance
analogous to $\bld{V}_{Y}$ that reflects the number of daily values
in each monthly average, plus a homoscedastic term reflecting the
fundamental discrepancy between AOD and true PM: $(\bld{V}_{A})_{ii}=\sigma_{A}^{2}+k(n_{i})\sigma_{\alpha}^{2}$. 

For simplicity and because I use monthly averages, I do not transform
the PM values, in contrast to other work on PM \citep{Smit:etal:2003,Yano:etal:2009},
but log or square root transformations are good alternatives. For
regression terms that represent sources, additivity on the original
scale makes more sense while for modifying variables such as meteorology,
log transformation to scale multiplicatively makes more sense. Achieving
both in a single model is not easily accomplished. I note that the
long right tail of CMAQ output may be accomodated through $\bld{\phi}$
because the CMAQ output is spatially correlated. As in \citet{Paci:etal:2009},
residuals from the various models indicated long-tailed behavior,
reasonably characterized by \textit{t} distributions with approximately
five degrees of freedom, albeit with right skew. Predictive performance
suggests that the influence of outliers is not extreme, but the use
of a \textit{t} distribution for the observation errors would be worth
exploring. However, this would be somewhat difficult to do in the
context of the additive error structure I derive above based on components
of variability in the PM measurements. 

I use several covariates calculated for individual cells at the four
km grid level: elevation at the cell centroids, population density,
and total length of roads in three road classes. Area PM emissions
from the 2002 EPA National Emissions Inventory (NEI) are calculated
as density of emissions per county, and the value for the county in
which a grid cell centroid falls is assigned to the grid cell. Population
density, road density, and area emissions are log-transformed to reduce
sparsity and pull in extremely large values in the right tail, and
I truncated the values of some outlying covariates to reduce extrapolation
problems. I used the NEI point source emissions strength and location
data in the flexible buffer modeling described in \citet[Appendix D]{Paci:Liu:2011},
creating a basis matrix that contributes columns to $\bld{Z}_{L}$.
This matrix leverages the additive structure of the mixed model representation
of a spline term to represent the effect of multiple point sources
at a given receptor location (i.e., a monitor or prediction point)
as the emission strength-weighted sum of a smooth distance effect
evaluated for each individual source-receptor pair. The distance function
is a universal function representing the effect of single source of
unit strength on a receptor as a function of the distance between
source and receptor and is estimated from the data. I calculate the
source strength-weighted sum of distance-weighted contributions from
fine PM primary source point emissions within a maximum distance (100
km) for each monitor, omitting sources emitting less than five tons
in 2002. For the proxy likelihood and for prediction on the grid,
I take a subgrid of 16 points within each grid box and calculate the
average sum of contributions from the point emissions within 100 km
of the points in the subgrid, as a simple approximation to the true
integral of the point emission effect over the grid cell. 

In general, the prior distributions for hyperparameters were non-informative,
with normal priors with large variances (and also lower and upper
bounds to prevent the MCMC from wandering in flat parts of the posterior)
for location parameters and uniform priors on the standard deviation
scale for scale parameters \citep{Gelm:2006}, with large upper bounds.
In all cases, the posterior distributions were much more peaked than
the prior distributions and away from the bounds, except for some
of the variance components for the coefficients of the regression
smooths, which I restricted to avoid overly wiggly smooth terms. Further
exploration of why these smooths tend toward less smooth functions
than expected scientifically and on whether simple linear relationships
would suffice, and might even improve out-of-sample prediction, would
be worthwhile.

I ran the MCMC for 10,000 iterations during the burn-in and 25,000
subsequently, retaining every 10th iteration to reduce storage costs.
I found reasonable convergence and mixing based on effective sample
size calculations and trace plots. I did not run multiple chains for
each month-validation set pair because of my use of multiple months
and validation sets, noting that predictive performance also helps
to justify the adequacy of the fitting.

\subsection*{Modifications for the eastern U.S. analysis}

Details here generally follow those just described, but for computational
efficiency, I represent $\bld{\phi}$ and $\bld{g}$ as TPS-MRFs on
the $73\times77=5621$ dimensional CMAQ grid over the eastern U.S.,
each with its own precision parameter. Covariate effects are represented
on the original four km base grid (now of dimension $669\times677=452,913$).
Pre-computation of $\bld{Z}_{A}$ in advance of the MCMC involves
the large matrix multiplication of a basis matrix and an averaging
matrix that represents the weighted average of four km cells within
each CMAQ pixel based on the amount of overlap. $\bld{Z}_{Y}$ also
represents the product of a mapping matrix and the original basis
matrices for the covariates. Given the relatively large sample size,
for this model I use 10 rather than five knots for each regression
smooth term.

The initial integration is over $\bld{\phi}^{*}=\{\bld{g},\bld{\phi}\}$
followed by integration over $\bld{b}$, which no longer includes
$\bld{b}_{g}$. The first integration is done by representing the
joint likelihood as 
\begin{eqnarray*}
\left(\begin{array}{c}
\bld{Y}\\
\bld{A}
\end{array}\right) & \sim & \mathcal{N}_{n_{Y}+n_{A}}\left(\left(\begin{array}{c}
\bld{Z}_{Y}\\
\bld{Z}_{A}
\end{array}\right)\bld{b}+\left(\begin{array}{cc}
\bld{P}_{Y} & \bld{0}\\
\beta_{1}\bld{P}_{A} & \bld{P}_{A}
\end{array}\right)\left(\begin{array}{c}
\bld{g}\\
\bld{\phi}
\end{array}\right)+\left(\begin{array}{c}
\bld{P}_{\delta}\\
\bld{0}
\end{array}\right)\bld{\delta},\left(\begin{array}{cc}
\bld{V}_{Y} & \bld{0}\\
\bld{0} & \bld{V}_{A}
\end{array}\right)\right),
\end{eqnarray*}
followed by analogous calculations to those in the previous section
to integrate over $\bld{b}$ and determine the marginal posterior
(up to the normalizing constant) for the remaining parameters and
the conditional normal posterior for $\bld{\delta}$ given the remaining
parameters and the data. Note that $\bld{P}_{Y}$ and $\bld{P}_{A}$
simply map from the CMAQ grid cells to the observations and CMAQ values. 

Some CMAQ pixels overlap four km cells entirely over water, and those
cells have undefined covariate values. I treat the CMAQ value in a
CMAQ pixel as reflecting the weighted average of $L(\cdot)$ from
only the land-based four km grid cells, with weights in $\bld{P}_{A}$
summing to one for each CMAQ pixel. I do not include CMAQ values in
the likelihood for CMAQ pixels with 60\% or more overlap with four
km cells that do not intersect land in the U.S.

For the point source emissions covariate, computational demands required
that I consider only point sources emitting more than 10 tons per
year within 50 km of a given location, with the integral approximation
using a subgrid with four, rather than 16, points within each four
km cell.

I ran the MCMC for 10,000 iterations during the burn-in and 20,000
subsequently, retaining every 10th iteration to reduce storage costs,
again finding reasonable convergence and mixing.
\end{document}